\documentclass{PoS}

\usepackage[table,dvipsnames]{xcolor}

\usepackage[official]{eurosym}
\usepackage{lineno}
\usepackage{url}

\newcommand\doi{{DOI: }}

\title{A next-generation ground-based wide field-of-view gamma-ray observatory in the southern hemisphere. }

\ShortTitle{SWGO: Southern Wide field-of-view Gamma-ray Observatory}

\author{\speaker{H. Schoorlemmer} on behalf of the SWGO collaboration\thanks{a list of the participating institutes can be found at  www.swgo.org}\\
        Max-Planck-Institut f\"ur Kernphysik, Heidelberg, Germany\\
        E-mail: \email{harmscho@mpi-hd-mpg.de}}

\abstract{The very-high-energy gamma-ray sky can be surveyed on a daily basis by particle-detector arrays at high (mountain) elevation. In the northern hemisphere the survey recently conducted by the HAWC gamma-ray observatory significantly enriched our knowledge about TeV gamma-ray sources. In this contribution, we will present an overview on the effort to realise a next-generation gamma-ray survey observatory in the southern hemisphere. We will discuss the unique science case for this observatory and how it will be embedded in the multi-messenger and multi-wavelength census of the non-thermal universe. In addition, we will introduce the newly founded international organisation that aims to realise this facility: The Southern Wide field-of-View Gamma-ray Observatory (SWGO) collaboration.}

\FullConference{36th International Cosmic Ray Conference -ICRC2019-\\
		July 24th - August 1st, 2019\\
		Madison, WI, U.S.A.}

\begin{document}

\section{Introduction}
\begin{figure}[!ht]
\begin{center}
\includegraphics[width=1\textwidth]{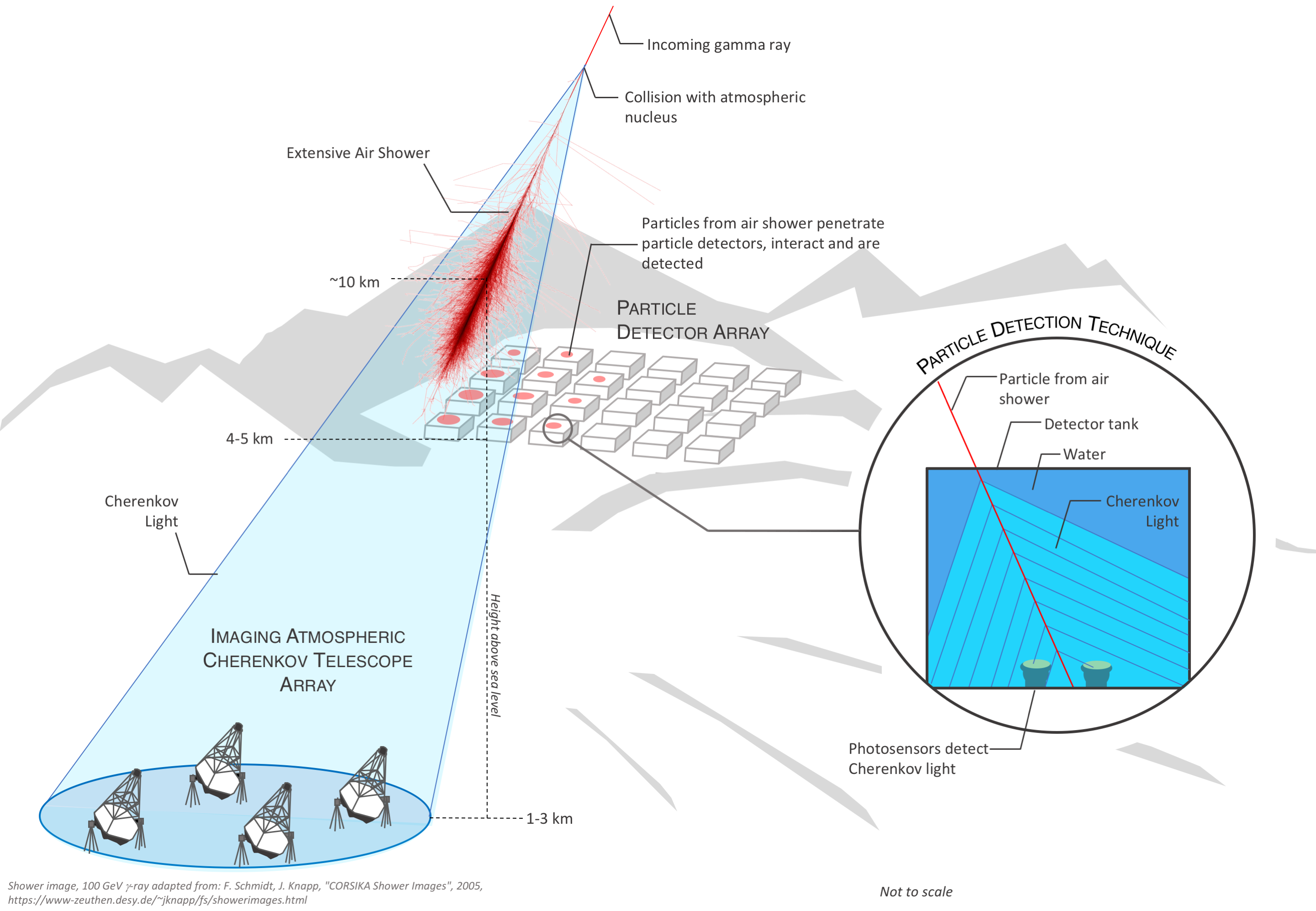}
\caption{Detection techniques for ground-based gamma-ray astronomy. The inset shows the working principle of a water-Cherenkov particle detector.}
\label{fig:concept}
\end{center}
\end{figure}

Ground-based gamma-ray astronomy relies on the detection of particle cascades initiated by the interaction of gamma rays with the Earth's atmosphere. Two techniques have been successfully applied to observe these cascades known as Extensive Air Showers (EASs): 
\begin{enumerate}
\item The detection of the Cherenkov light emitted by the relativistic EAS with optical Imaging Atmospheric Cherenkov Telescopes (IACTs).
\item The detection of the cascade particles with an array of particle detectors at a high altitude site. 
\end{enumerate}
Figure \ref{fig:concept} illustrates these two techniques. They complement each other in the way they observe the gamma-ray sky. On the one hand, IACTs are pointing instruments with a typical field-of-view limited to a few degrees and a duty-cycle of 10-20\%, while particle detector arrays have a very wide field-of-view ($\sim90^\circ$) and a roughly 100\% duty-cycle. On the other hand, since particle detector arrays rely on the EAS particles arriving at ground level, the energy threshold is higher and the accuracy with which the energy and direction of the primary gamma ray can be determined is typically less than with IACTs which observe the development of the EAS in the atmosphere.

In the northern hemisphere, the High Altitude Water Cherenkov Gamma-ray Observatory (HAWC~\cite{2017ApJ...843...39A}) has been observing the sky over the last few years, and recently the Large High Altitude Air Shower Observatory (LHAASO~\cite{ScienceLHAASO}) has started operations. Both of these observatories use (mainly) the water-Cherenkov detectors technique to measure the air shower particles at ground level and will continue to survey the northern sky in the coming years. However, no such facility has ever been operated in the southern hemisphere, leaving a key part of the gamma-ray sky unobserved by this technique. An international collaboration of scientists recently joined forces in order to develop a detailed plan for a next-generation facility of this type in the southern hemisphere under the name Southern Wide field-of-view Gamma-ray Observatory (SWGO). In this contribution, the science case for the observatory is briefly summarised and we introduce the recently founded collaboration and its plans.

\section{Science with SWGO}\label{science}
\begin{figure}[!ht]
\begin{center}
\includegraphics[width=.99\textwidth]{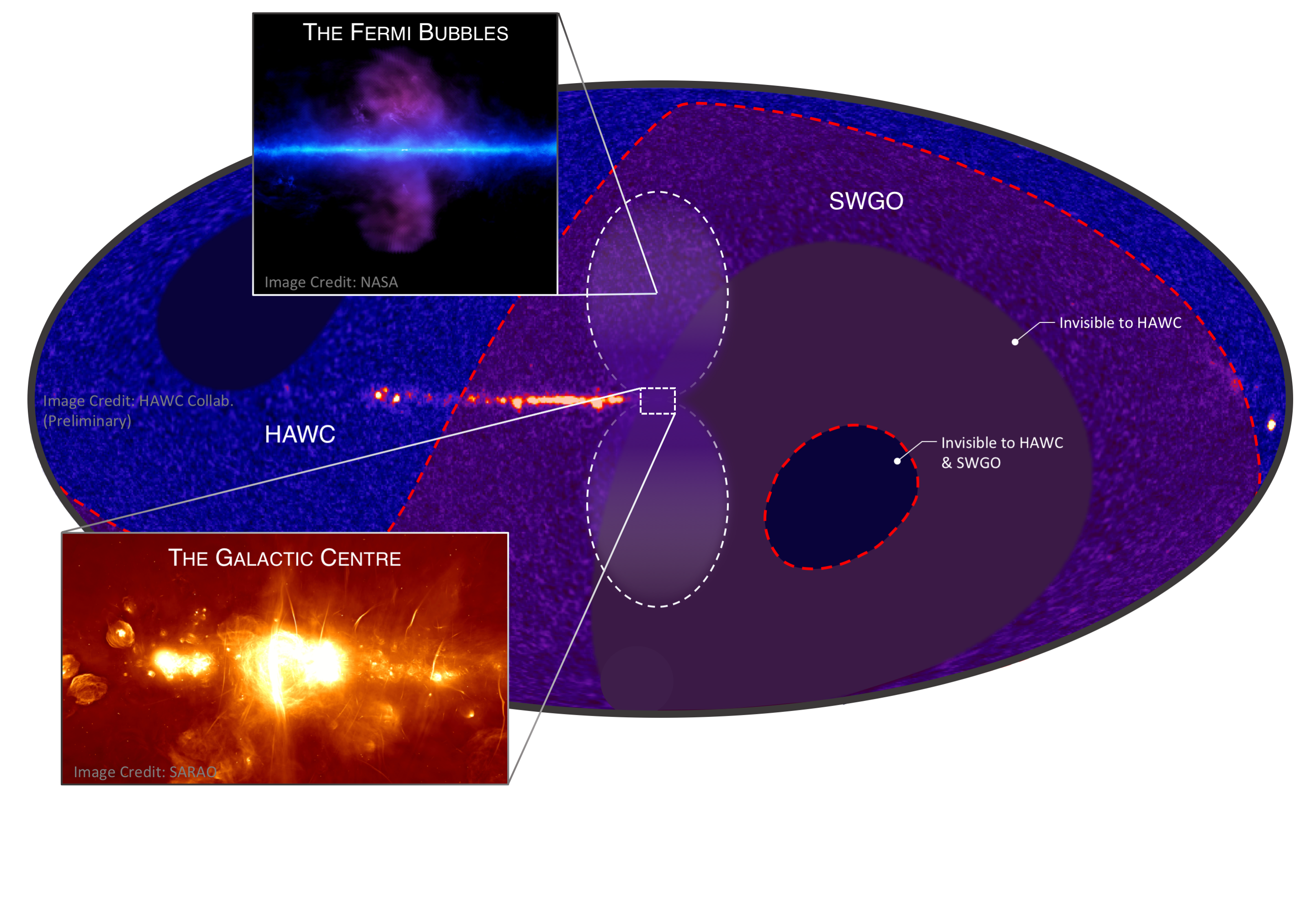}
\caption{Sky coverage of the Southern Wide field-of-view Gamma-ray Observatory overlaid on the survey carried out by HAWC in the northern hemisphere. The HAWC survey has so far identified more than 50 sources. LHAASO will have roughly the same sky coverage as HAWC.}
\label{fig:sky}
\end{center}
\end{figure}

The observable sky for SWGO is illustrated in Figure~\ref{fig:sky} together with some of the prime target astrophysical sources.

The scientific case of this future facility has been recently explored in great detail, which resulted in a scientific `white paper' \cite{ScienceCase} supported by a large number of scientists ($>$100).  Here we will give a brief overview of the science case, but refer the reader for more details to \cite{ScienceCase} and the dedicated contribution at this conference \cite{ScienceICRC}. To perform different science case studies, the response of an observatory was simulated with performance parameters from HAWC \cite{2017ApJ...843...39A} that were scaled by the use of MC simulations of EASs to a significantly larger array at a higher altitude of 5\,km above sea level. 
 
Compared to the southern IACTs, the current-generation High Energy Stereoscopic System (H.E.S.S.) and the future Cherenkov Telescope Array (CTA~\cite{2019scta.book.....C}), SWGO has the potential to exceed their point-source sensitivity at the high-energy part ($\sim 30$\,TeV)of the gamma-ray spectrum. This is a direct consequence of its $\sim$100\% up-time and large effective area. This will enable a deep unbiased survey of a large fraction of the Galactic plane to search for (very) hard spectra gamma-ray sources. This survey will lead to the identification of candidate astrophysical sources that can accelerate particles up to PeV energies. The observations of SWGO, in combination with follow-up and contemporary observations by CTA and neutrino telescopes such as KM3Net and IceCube, will provide  together unprecedented insight into the physics of cosmic particle accelerators.

The point-source sensitivity is a way to compare observatories, but does not convey the full scientific potential, especially the measurement of (very) extended emission regions is not captured well by this figure of merit. Mapping of gamma-ray regions that extend over more than several degrees on the sky is one of the main objectives for SWGO. Its wide field-of-view will provide for an extremely accurate determination of the background (typically $<10^{-4}$), which is crucial to observe faint emission extended over more than multiple degree angular scales. The southern sky contains the most interesting known extended gamma-ray sources, namely the Galactic Centre region and the Fermi Bubbles. In addition, there is high potential to discover `halos' of gamma-ray emission around evolved sources, as HAWC did for two nearby pulsars \cite{2017Sci...358..911A}. These are systems that provide a unique probe of electron propagation. Such sources the ones in the vicinity of the Earth will be very extended and likely to appear at high Galactic latitude which makes them an ideal target for surveying instruments such as SWGO.

The combination of good sensitivity for extended and high-energy sources will in particular benefit the hunt for a potential gamma-ray signal of annihilating dark matter accumulated in the inner part of our own galaxy. It is expected that the sensitivity in terms of the average weighted  cross-section will be at a similar level at that reached by CTA, but shifted towards higher mass dark-matter particles \cite{2019arXiv190603353V}. Therefore, the dark matter mass reach over which a thermal-relic cross section can be ruled will be expanded to a higher mass range.
The high up-time and wide field-of-view also makes SWGO a monitoring device for variable and transient astrophysical phenomena. Most of the known TeV variable and transient phenomena are of extra-galactic origin, such as gamma-ray bursts (GRBs) and flaring active galactic nuclei (AGNs). Since the mean free path of gamma rays decrease significantly with energy above hundreds of GeV, due to interactions with intergalactic radiation fields, the sub-TeV sensitivity of SWGO is of paramount importance. This will partially be reached by deploying it at a high altitude site, but  also by increasing the size of the array and improving the  sensitivity of individual detection units with respect to HAWC. With the anticipated improvement for the `straw man' design used in the science case white paper \cite{ScienceCase}, SWGO will be able to detect (bright) GRBs and monitor the activity of tens of AGNs and issue alerts to be followed up by other observatories. One other advantage of SWGO is that it will be possible to look into archival data and search for potential gamma-ray counterparts for observations made by other contemporary observatories on all time and angular scales. 

The combination of HAWC and/or LHAASO data with cosmic-ray observations from SWGO will permit a nearly all-sky observation of the cosmic-ray anisotropy, connecting the propagation of cosmic rays to the structure of the local heliospheric and interstellar magnetic fields~\cite{2019ApJ...871...96A}.
In the immediate solar neighbourhood, SWGO will play a primary role in investigating the production of VHE gamma rays produced in cosmic-ray interactions with the solar atmosphere~\cite{2019BAAS...51c.194N}, since few existing and planned TeV gamma-ray detectors can monitor the Sun.
Observations of the Sun by SWGO will constrain the production mechanism of these gamma rays and its relationship to the solar cycle, which are currently unknown~\cite{2011ApJ...734..116A,2017PhRvD..96b3015Z}.
Finally, measurements of the ``shadow'' in the flux of Galactic cosmic rays created by the Sun will provide an independent probe of the magnetic fields near the Sun and in the interplanetary environment~\cite{2013PhRvL.111a1101A}.

\section{Research and Development collaboration}

\begin{figure}
\begin{center}
\includegraphics[width=1\textwidth]{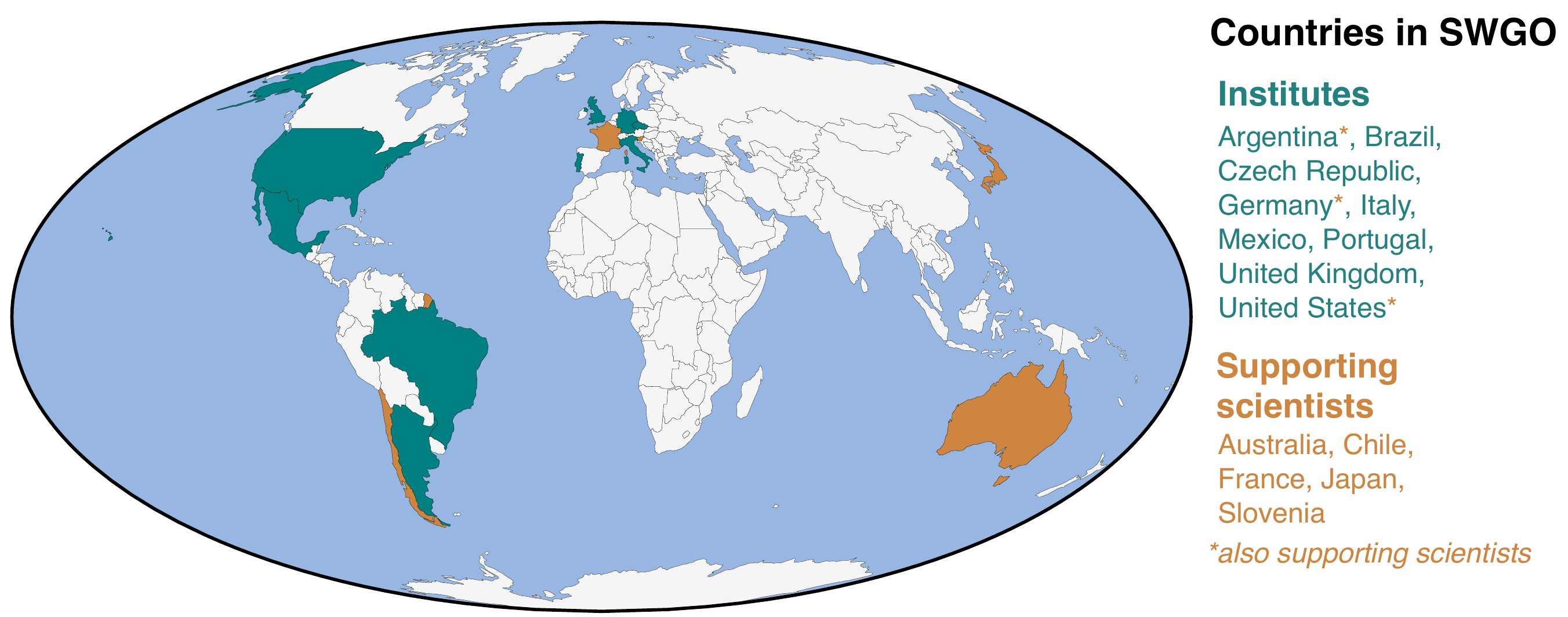}
\caption{The countries involved in the international SWGO R\&D collaboration. In total, the collaboration has already over more than one hundred members.}
\label{fig:collab}
\end{center}
\end{figure}

By signing a statement of interest (SoI) on July 1st 2019, 41 institutions from nine countries have formed a collaboration with the intent to pursue research and development towards SWGO (see \href{https://www.swgo.org/SWGOWiki/doku.php?id=collaboration}{www.swgo.org}  for more information and Figure \ref{fig:collab}). This new collaboration is a step forward beyond the existing SGSO alliance, LATTES\cite{LATTES} and STACEX projects, and the initial duration of the R\&D programme is agreed to be three years. The objective is to have a single full design and proposal at the end of the R\&D programme. Individual scientists not affiliated with a member institution of the SWGO collaboration can on request become supporting scientists, welcome to join any of the working groups, with full access to programme-internal information. The execution of the R\&D program is overseen by a steering committee consisting of national representatives.
The following working groups are being established:
\begin{itemize}
    \item Science case development
    \item Simulations, analysis and array optimisation
    \item Candidate site evaluation
    \item Detector design and development
    \item Outreach
\end{itemize}

The concept for the future observatory is as follows:
A gamma-ray observatory based on ground-level particle detection, with close to 100\% duty cycle and order steradian field-of-view, located in South America at a latitude from -30 to -10 degrees at an altitude of 4.4\,km or higher. It will  cover an energy range from 100s of GeV to 100s of TeV. The detection technique will be based primarily on water Cherenkov detector units,  with a high fill-factor core detector with area considerably larger than HAWC and significantly better sensitivity and a low density outer array.
There will be the  possibility of extensions and/or enhancements. The design will be modular and scalable and the observatory will aim for a target cost of \euro{}40-50M. The effort will be organised as a collaboration, constructing the instrument and exploiting the scientific data together.  

\subsection{Software development for array simulations}
Currently a common software framework is under development, hosted for collaboration members on \url{www.gitlab.com} and will be publicly available in the near future. This framework is being developed based upon that of HAWC to speed up the development phase and compare performance.    

\begin{figure}
\begin{center}
\includegraphics[width=.99\textwidth]{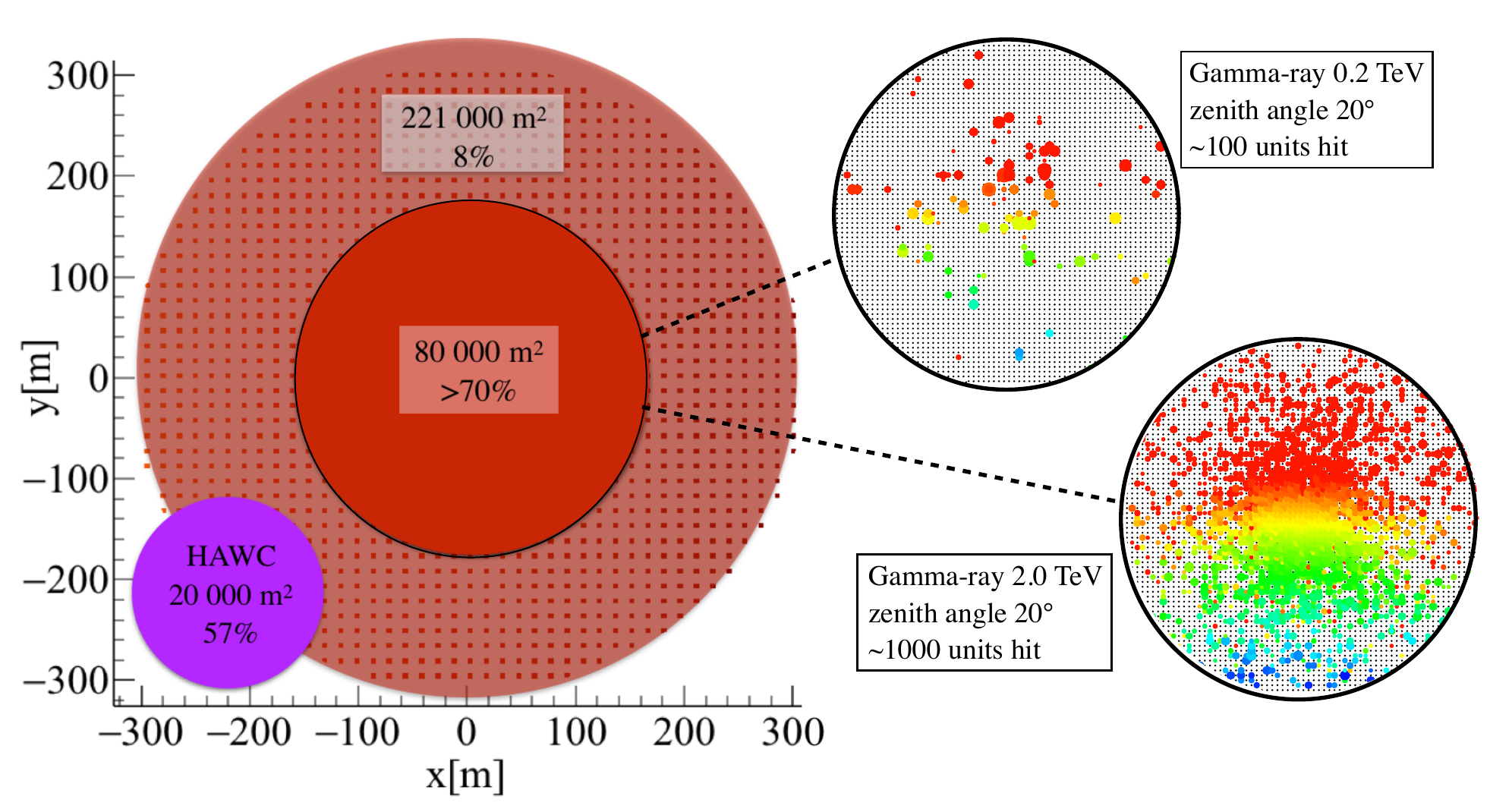}
\caption{A conceptual design of an array at 5\,km altitude and its simulated response to two example gamma-ray events. The colour coding indicates the time gradient of the signals recorded by the detection units.}
\label{fig:example}
\end{center}
\end{figure}

The software has been adapted in such a way that different detection unit designs can be compared easily in several array geometries. For more information see \cite{OptWCD_ICRC} at this conference.  As an example we show the response of a conceptual array of two gamma-ray events. The 200\,GeV gamma-ray example illustrates the large spread of particles which is typical for these sub-TeV air showers \cite{Schoorlemmer2019}. This emphasises the need for a large array with high ground coverage in order to optimise the low-energy response of the observatory. The 2 TeV example shows the potential to reconstruct at this intermediate energy the properties of the gamma rays with high accuracy.  

\subsection{Site Search}
In recent years several site locations have been identified that fulfil the boundary criteria on altitude and latitude of the observatory. These sites are located in Argentina, Bolivia, Chile, and Peru and have been visited by individual members of the collaboration. The goal is to characterise these sites in detail such that at the end of the three years R\&D phase a decision can be made on the final location of the observatory. There are several aspects to consider that will influence the observatory design, such as the annual temperature profile, the vicinity of nearby water resources, and availability of existing infrastructure. 

\section{Conclusion}
The recent survey of the TeV gamma-ray sky in the northern hemisphere conducted by the HAWC observatory has revealed many interesting new sources and both LHAASO and HAWC will continue this survey in the coming years. In the southern hemisphere, which hosts many interesting astrophysical targets, a new observatory is being planned to reach almost full sky coverage with this technique and a new collaboration was created to organise this effort. This facility will be in close scientific coordination with CTA, recognising the synergy and complementarity between these instruments. It will monitor the southern sky for transient phenomena and be embedded in the global multi-messenger and multi-wavelength network. 
With its wide field-of-view it will map out extended regions of emission with unprecedented sensitivity. 

\bibliographystyle{new-etal}
\bibliography{references}

\begin{thebibliography}{10}
\input{babelbst.tex}
\newcommand{\Capitalize}[1]{\uppercase{#1}}
\newcommand{\capitalize}[1]{\expandafter\Capitalize#1}

\bibitem{2017ApJ...843...39A}
A.~U. {Abeysekara}, \bbletal{}, \emph{{Observation of the Crab Nebula with the
  HAWC Gamma-Ray Observatory}}, \apj, \textbf{843}, 39 (2017),
  \doi{10.3847/1538-4357/aa7555}

\bibitem{ScienceLHAASO}
X.~{Bai}, \bbletal{}, \emph{{The Large High Altitude Air Shower Observatory
  (LHAASO) Science White Paper}}, arXiv e-prints, arXiv:1905.02773 (2019)

\bibitem{ScienceCase}
A.~{Albert}, \bbletal{}, \emph{{Science Case for a Wide Field-of-View
  Very-High-Energy Gamma-Ray Observatory in the Southern Hemisphere}}, arXiv
  e-prints, arXiv:1902.08429 (2019)

\bibitem{ScienceICRC}
F.~{Sch\"ussler}, \emph{{Science Case for a Wide Field-of-View Very-High-Energy
  Gamma-Ray Observatory in the Southern Hemisphere}}, \bblin{}
  \emph{Proceedings of Science: PoS(ICRC2019)786} (2019)

\bibitem{2019scta.book.....C}
{Cherenkov Telescope Array Consortium}, \bbletal{}, \emph{{Science with the
  Cherenkov Telescope Array}}, World Scientific Publishing Co. Pte. Ltd.
  (2019), \doi{10.1142/10986}

\bibitem{2017Sci...358..911A}
A.~U. {Abeysekara}, \bbletal{}, \emph{{Extended gamma-ray sources around
  pulsars constrain the origin of the positron flux at Earth}}, Science,
  \textbf{358}, 911 (2017), \doi{10.1126/science.aan4880}

\bibitem{2019arXiv190603353V}
A.~{Viana}, \bbletal{}, \emph{{Searching for Dark Matter in the Galactic Halo
  with a Wide Field-of-View TeV Gamma-ray Observatory in the Southern
  Hemisphere}}, arXiv e-prints, arXiv:1906.03353 (2019)

\bibitem{2019ApJ...871...96A}
A.~U. {Abeysekara}, \bbletal{}, \emph{{All-sky Measurement of the Anisotropy of
  Cosmic Rays at 10 TeV and Mapping of the Local Interstellar Magnetic Field}},
  \apj, \textbf{871}, 96 (2019), \doi{10.3847/1538-4357/aaf5cc}

\bibitem{2019BAAS...51c.194N}
M.~{Nisa}, \bbletal{}, \emph{{The Sun at GeV-TeV Energies: A New Laboratory for
  Astroparticle Physics}}, \bblin{} \emph{\baas}, \bblvol{}~51, 194 (2019)

\bibitem{2011ApJ...734..116A}
A.~A. {Abdo}, \bbletal{}, \emph{{Fermi Large Area Telescope Observations of Two
  Gamma-Ray Emission Components from the Quiescent Sun}}, \apj, \textbf{734},
  116 (2011), \doi{10.1088/0004-637X/734/2/116}

\bibitem{2017PhRvD..96b3015Z}
B.~{Zhou}, \bbletal{}, \emph{{TeV solar gamma rays from cosmic-ray
  interactions}}, \prd, \textbf{96}, 023015 (2017),
  \doi{10.1103/PhysRevD.96.023015}

\bibitem{2013PhRvL.111a1101A}
M.~{Amenomori}, \bbletal{}, \emph{{Probe of the Solar Magnetic Field Using the
  ``Cosmic-Ray Shadow'' of the Sun}}, \prl, \textbf{111}, 011101 (2013),
  \doi{10.1103/PhysRevLett.111.011101}

\bibitem{LATTES}
P.~{Assis}, \bbletal{}, \emph{{Design and expected performance of a novel
  hybrid detector for very-high-energy gamma-ray astrophysics}}, Astroparticle
  Physics, \textbf{99}, 34 (2018), \doi{10.1016/j.astropartphys.2018.02.004}

\bibitem{OptWCD_ICRC}
S.~{Kunwar}, H.~{Schoorlemmer}, \& J.~{Hinton}, \emph{{Optimization studies of
  a Water Cherenkov Detector for Gamma-ray Astronomy}}, \bblin{}
  \emph{Proceedings of Science: PoS(ICRC2019)720} (2019)

\bibitem{Schoorlemmer2019}
H.~Schoorlemmer, J.~Hinton, \& R.~L{\'o}pez-Coto, \emph{Characteristics of
  extensive air showers around the energy threshold for ground-particle-based
  $\gamma$-ray observatories}, The European Physical Journal C, \textbf{79},
  427 (2019), \doi{10.1140/epjc/s10052-019-6942-x}

\end{thebibliography}

\end{document}